\documentclass[reprint,amsmath,amssymb,aps,prb]{revtex4-2}
\usepackage{graphicx}
\usepackage{dcolumn}
\usepackage{bm}
\usepackage{hyperref}
\usepackage{epstopdf}
\usepackage{adjustbox}
\usepackage{float}
\usepackage{cancel}
\usepackage[normalem]{ulem}
\usepackage{xcolor}

\usepackage{amsmath}
\usepackage{mathrsfs}

\begin{document}

\preprint{APS/123-QED}

\title[mode = title]{Thermal effect on microwave pulse driven magnetization switching of Stoner particle}

\author{S. Chowdhury}
\affiliation{Physics Discipline, Khulna University, Khulna 9208, Bangladesh}
\author{M. A. S. Akanda}
\affiliation{Physics Discipline, Khulna University, Khulna 9208, Bangladesh}
\author{M. A. J. Pikul}
\affiliation{Department of Physics, Colorado State University, Fort Collins, Colorado 80523, USA}
\author{M. T. Islam}
\email[Correspondence email address:]{torikul@phy.ku.ac.bd}
\affiliation{Physics Discipline, Khulna University, Khulna 9208, Bangladesh}
\affiliation{Center for Spintronics and Quantum Syetems, State Key Laboratory for Mechanical Behavior of Materials,  Xi'an Jiaotong University, No. 28 Xianning West Road Xi'an, Shanxi 710049,  China}
\author{Tai Min}
\affiliation{Center for Spintronics and Quantum Syetems, State Key Laboratory for Mechanical Behavior of Materials,  Xi'an Jiaotong University, No. 28 Xianning West Road Xi'an, Shanxi 710049,  China}

\begin{abstract}
Recently it has been demonstrated that the cosine chirp microwave pulse (CCMP) is capable of achieving fast and energy-efficient magnetization-reversal of a nanoparticle with zero-Temperature. However, we investigate the finite temperature, $T$ effect on the CCMP-driven magnetization reversal using the framework of the stochastic Landau Lifshitz Gilbert equation.  At finite Temperature, we obtain the CCMP-driven fast and energy-efficient reversal and hence estimate the maximal temperature, $T_{max}$ at which the magnetization reversal is valid. $T_{max}$ increases with increasing the nanoparticle cross-sectional area/shape anisotropy up to a certain value, and afterward $T_{max}$  decreases with the further increment of nanoparticle cross-sectional area/shape anisotropy. This is because of demagnetization/shape anisotropy field opposes the magnetocrystalline anisotropy, i.e., reduces the energy barrier which separates the two stable states. For smaller cross-sectional area/shape anisotropy, the controlling parameters of CCMP show decreasing trend with temperature. We also find that with the increment easy-plane shape-anisotropy, the required initial frequency of CCMP significantly reduces. For the larger volume of nanoparticles, the parameters of CCMP remains constant for a wide range of temperature which are desired for the device application. Therefore, The above findings might be useful  to
realize the CCMP-driven fast and energy-efficient magnetization reversal in realistic conditions.

\end{abstract}
\maketitle

\section{Introduction}
Obtaining swift and efficient magnetization switching of a single nanoparticle has attracted much attention because of its non-volatility \cite{sun2000,woods2001,zitoun2002} and speedy data processing properties \cite{hillebrands2003}. In the last decades, several controlling parameters, for instance, magnetic fields, microwaves fields \cite{hubert1998,sun2005}, spin-polarized electric current and spin-orbit torque \cite{slonczewski1996,berger1996,tsoi1998,Katine2000,Waintal2000,sun2000a,suns2003,stiles2002,bazaliys2004,koch2004,wetzels2006,manchon2008,miron2010,miron2011,liu2012} are employed to reverse magnetization reversal.  However, these methods are facing specific challenging issues in memory-device applications. Particularly, in the case of the magnetic field, it requires a large field and a longer switching time \cite{hubert1998}; also field localization to a bit-cell is a bottleneck. On the other hand, the spin-polarized-current can induce magnetization switching by the mechanism of spin transfer torque (STT) and$\slash$or spin-orbit torque (SOT) \cite{zhang2018breaking,vlasov2022optimal, aryal2022magnetic, wang2021field, krizakova2022spin, ovalle2023spin}. However, for STT-MRAM-based devices, the threshold current density is large and it creates Joule heat which, in turn, causes the device lifetime and reliability issues \cite{grollier2003, morise2005, taniguchi2008, SUZUKI200993, zsun2006, wang2007, wange2008}. For SOT-MRAM-based device applications, the main hindrance is that the requirement of two transistors for each bit-cell enhances the area of the bit-cell \cite{kim2014multilevel}. Later on, people employ the microwave field with constant or time-dependent frequency profile to drive magnetization switching at zero temperature \cite{bertotti2001, sunz2006, denisov2006, okamoto2008, zhu2010, thirion2003, rivkin2006, wangc2009, barros2011, barrose2013, tanaka2013, klughertzx2014,juthy2022shape,islam2018subnanosecond}. However, without considering the thermal effect, the recent study \cite{islam2021fast} reported that the swift and energy-efficient magnetization switching is obtained by a cosine chirped microwave pulse (cosine CMP). This is because the frequency changing (of Cosine CMP) is closely matched with the magnetization precession frequency which leads to the stimulated energy absorption (emission) by magnetization efficiently before (after) crossing the energy barrier.  However, in practice, the temperature is prevailing everywhere, and devices are operated at room temperature. So, from the practical point of view, it is interesting to verify whether the Cosine CMP still efficiently drives magnetization switching at finite temperature. In addition, the study  \cite{islam2021fast} reported that the increase of easy-plane shape anisotropy (i.e., the increase of demagnetization field) makes the magnetization switching faster which is expected for device application. But, the increase of the demagnetization field (which opposes the magnetocrystalline anisotropy) reduces the height of the energy barrier, originated by anisotropy, which may cause thermal instability issues at room$\slash$operating temperature.
Thus at operating temperature, there is a possibility of spontaneous magnetization switching which may increase the error rate, which is actually undesired. Therefore, in this study, we include the finite/room temperature in the system and relax it, afterward we investigate the cosine CMP-driven magnetization switching to check whether the switching is robust at operating (room) temperature and how the parameters (i.e., the optimal initial frequency and field amplitude) of cosine CMP are altered with temperature.


This study interestingly finds that the cosine CMP-driven swift and energy-efficient switching is robust even with a finite$\slash$room temperature. For the considered nanoparticles$\slash$stoner particle, we also estimate the maximal temperature, $T_\text{m}$ at which the magnetization is valid. $T_\text{m}$ increases with increasing the nanoparticle volume (by increasing cross-sectional area, $A=xy$) or shape-anisotropy coefficient up to a certain value, and $T_\text{m}$ significantly decreases with the further increment of nanoparticle cross-sectional area$\slash$shape-anisotropy coefficient. Here the demagnetization$\slash$shape anisotropy reduces the effective uniaxial anisotropy i.e., reduces the energy barrier which separates the two stable states. Still, the cosine CMP-driven magnetization switching is valid for a wide operating temperature above room temperature. For the smaller volume of nanoparticles$\slash$stoner particle, the controlling parameters of cosine CMP, i.e., the minimal amplitude $H_\text{mw}$ (T), the optimal rate $R$ (GHz), and the minimal initial frequency $f_0$ (GHz) show a decreasing trend with temperature. We also find that with the increment of easy-plane shape-anisotropy, the required initial frequency of cosine CMP significantly reduces. For the larger volume of nanoparticles, the parameters of cosine CMP remain constant for a wide range of temperature that is desired for the device application. Note that this strategy might be employed to switch the magnetization of other materials, for instance, synthetic antiferromagnetic/ ferrimagnetic nanoparticles by in-plane cosine chirped current pulse via spin-orbit torque. Therefore,   the above findings might be useful to realize the cosine CMP-driven fast and energy-efficient magnetization switching in practical spintronics device applications with considering realistic conditions.

\section{Model and Method}
\begin{figure}
     \centering
     \includegraphics[width=85mm]{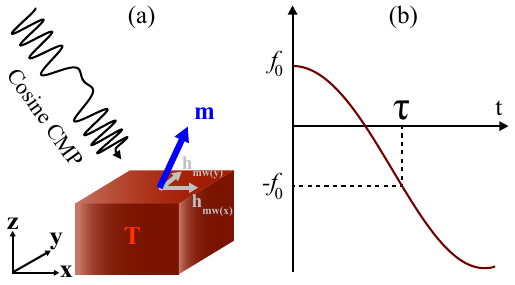}
     \caption{(a) Schematic figure shows a single domain nanoparticle$\slash$stoner particle with in-plane orthogonal microwave field components at finite temperature, where $\mathbf{m}$ denotes the direction of magnetization. (b) Time-depended frequency outline for cosine CMP from $+f_0$ to $-f_0$.}
     \label{figure:1}
 \end{figure}

We assume a single-domain magnetic nanoparticle/Stoner particle of volume $V=Ah$, where $A$ is the cross-sectional area and $h$ is thickness, and its uniaxial anisotropy is directed along $z$ axis at temperature $T$, as shown in Figure \ref{figure:1}(a). The nanoparticle's size is chosen in such a way that the magnetization of can be treated as a macrospin which is indicated by the unit-vector $\mathbf{m}$ with saturation magnetization $M_s$. An easy-plane shape anisotropy is approximated the demagnetization field which is opposite to the magnetocrystalline anisotropy field.
The shape anisotropy field is indicated as $\mathbf{H}_\text{shape} = K_\text{shape} m_z \hat{\mathbf{z}}$, where $K_\text{shape} = -\mu_0(N_z-N_x)M_s$ is  the shape anisotropy coefficient,  $N_z$ and $N_x$ are demagnetization factors \cite{dubowik1996,aharoni1998} and $\mu_0 = 4\pi \times 10^{-7} \text{ N}/\text{A}^2$ is the vacuum magnetic permeability. The total anisotropy is dominated by the magnetocrystalline anisotropy $\mathbf{H}_\text{ani} = K_\text{ani} m_z \hat{\mathbf{z}}$ and hence the magnetic particle possesses two stable states (i.e., the $\mathbf{m}$ aligns parallel to $\hat{\mathbf{z}}$ and $-\hat{\mathbf{z}}$).

At finite temperatures, the dynamics of magnetization are governed by the stochastic Landau Lifshitz Gilbert (sLLG) with the application of a circularly polarized cosine CMP \cite{gilbert2004}.
\begin{align}
    \frac{d\mathbf{m}}{d t} = - \gamma \mathbf{m} \times [\mathbf{H}_{\text{tot}} +{\mathbf{H}_\text{th}}] + \alpha \mathbf{m} \times \frac{d\mathbf{m}}{d t},
\end{align}
where  $\alpha$ and $\gamma$ are the Gilbert damping constant and the gyromagnetic ratio respectively. Additionally, the total magnetic field ($\mathbf{H}_\text{tot}$) consists of the effective field, which comes from the exchange field $\frac{2 A_\text{ex} }{M_s}\nabla^2 \mathbf{m}$, and effective the easy-axis anisotropy field along $z$ direction, and the external microwave field ($\mathbf{H}_\text{mw}$).

The stochastic thermal field is denoted by $\mathbf{H}_\text{th}$, which comes from a finite temperature. The thermal field exhibits the following relations which are described by the Gaussian process \cite{fuller1963thermal, Nowak2008}.
\begin{equation}
    \begin{gathered}
        \langle H_{\text{th},ip}(t)\rangle = 0,\\
        \langle H_{\text{th},ip}(t)H_{\text{th},jq}(t+\Delta t)\rangle = \frac{2\alpha k_{\text{B}}T} {\gamma  M_s \Delta V}\delta_{ij}\delta_{pq}\delta (\Delta t),
    \end{gathered}
\end{equation}

where $k_{B}$ is the Boltzman constant, $p$ and $q$ are the Cartesian thermal field components, $\Delta V$ is the volume of a single micromagnetic cell, and $i$ and $j$ designate the micromagnetic cells. Depending on the temperature, the thermal$\slash$random field is generated as 
\begin{equation}
    \mathbf{H}_{\text{th, i,p}} = \Vec{\eta} \sqrt{\dfrac{2\alpha k_\text{B}T}{\gamma M_\text{s}\Delta V\Delta t}}
\end{equation}
where $\Delta t$ is the time step and $\Vec{\eta}$ is a random vector which changes with time step and provides the normal distribution with zero average.

Without microwave field and with zero-temperature,  there are two ground states (or energy minima) $\mathbf{m} \parallel \hat{z}$ and $\mathbf{m}\parallel -\hat{z}$ in which the magnetization likes to stay due to uniaxial anisotropy. The main task is to switch the magnetization from one energy minima to another energy minima, purposely the study \cite{islam2021fast} reported that, at zero temperature,  the cosine CMP is efficient to achieve fast switching. The cosine CMP is constructed as $\mathbf{H}_\text{mw} = H_\text{mw} \left[ \cos\phi(t) \hat{\mathbf{x}} + \sin\phi(t) \hat{\mathbf{y}}\right]$,
where $H_\text{mw}$ is microwave amplitude and $\phi(t)$ is the phase.  $\phi(t)$ is denoted as $2 \pi f_0 \cos \left(2 \pi R t \right) t$, $R$ (in GHz) is the controlling parameter, and $f(t)$ is the instantaneous frequency of cosine CMP, is defined as $f(t) =\frac{1}{2\pi} \frac{d\phi}{dt}= f_0 \left[\cos \left(2 \pi Rt \right) - \left(2 \pi Rt\right) \sin \left(2 \pi Rt\right) \right]$ which sweeps from $+f_0$ to final $-f_0$  as shown in Figure \ref{figure:1}(b). The physical picture of obtaining the fast and energy-efficient magnetization switching is that the cosine CMP induces stimulated microwave absorption (emissions) by (from) the macro-spin before (after) crossing over the energy barrier at zero temperature and, for detail, the formulation of the rate of energy change is given in the appendix \cite{islam2021fast}. 
In this study, we use the material parameters reported in Ref. \cite{lu2009m}, $M_{s}=10^{6}$ A$\slash$m, ${H_\text{k}}$ = $0.75$ T or $3.75 \times10^{5}$ J$\slash$m$^{3}$, $\gamma = 1.76\times 10^{11} \text{ rad}/(\text{T}\cdot\text{s})$, exchange constant  ${A_\text{ex}}$  = $13 \times 10^{-12}$ J$\slash$m, and $\alpha= 0.01$ to mimic Permalloy. This study shows a strategy that would work for other materials also since Permalloy does not possess such a high anisotropy \cite{Hasegawa2017, dieny2017perpendicular}.  MuMax3 Package \cite{vansteenkiste2014} has been used to solve the stochastic-LLG equation using the adaptive-Heun solver. 

We study the nanoparticles by increasing the cross-sectional area $A=xy$ while the thickness $h = 8$ nm is kept constant. Particularly, the nanoparticles with the areas are $A_1=8\times8$ nm$^2$, $A_2=12\times12$ nm$^2$, $A_3 =16\times16$ nm$^2$, and $A_4=22\times22$ nm$^2$ so that the demagnetization field is induced in the opposite direction of uniaxial anisotropy. We discretize the sample by the unit-cell size $2\times2\times2$ nm$^3$. For efficient and stable calculation \cite{islam2019,islam2023role}, we employ the fixed time step of $10^{-14}$ s. According to the practical requirement, we consider the switched state is obtained if the magnetization reaches at $m_z=-0.7$.

\begin{table*}
    \caption{\label{table-1}Optimal rate $R$, Minimal initial frequency $f_0$, and Minimal amplitude $H_\text{mw}$ in different temperature.}
    \begin{tabular}{@{}|l|l|l|l|l|l|}
        \hline
        Cross-sectional & Shape & Temperature, & Optimal $R$ & Minimal initial & Minimal\\
        area, & anisotropy, & $T$ (K) & rate, & frequency, & Amplitude,\\
        $A$ (nm$^2$) & ${K}_\text{shape}$ (T) &  & $R$ (GHz) & $f_0$ (GHz) & $H_\text{mw}$ (T)\\
        \hline
        $A_1$ & 0 & $T$= 0 & 17.22 & 18.8 & 0.0450 \\
        && $T$= 300 & 15.12 & 18.8 & 0.0445 \\
        && $T$= 1000 & 16.38 & 18.7 & 0.0439 \\
        \hline
        $A_2$ & 0.17718 & $T$= 0 & 17.64 & 17.7 & 0.0450 \\
        && $T$= 300 & 16.38 & 17.7 & 0.0450 \\
        && $T$= 1400 & 14.28 & 17.6 & 0.0447 \\
        \hline
        $A_3$ &0.3064 & $T$= 0 & 17.22 & 13.5 & 0.0450 \\
        && $T$= 300 & 16.8 & 13.1 & 0.0450 \\
        && $T$= 1500 & 15.96 & 13.5 & 0.0447 \\
        \hline
        $A_4$ &0.4459 & $T$= 0 & 27.3 & 7.7 & 0.0450 \\
        && $T$= 300 & 27.3 & 7.1 & 0.0450 \\
        && $T$= 800 & 27.3 & 6.3 & 0.0450 \\
        \hline
    \end{tabular}
\end{table*}

\section{Numerical Results}

\begin{figure}[b]
    \centering
    \includegraphics[width=0.45\textwidth]{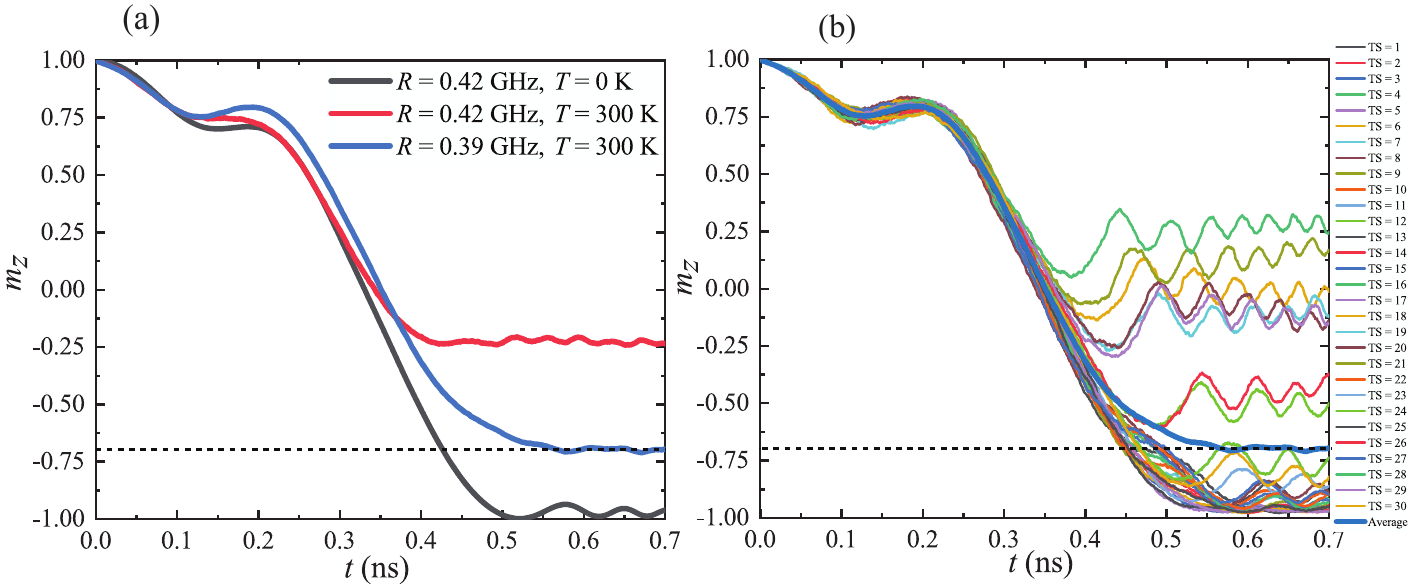}
    \caption{(a) The time evolutions of $m_z$ of nanoparticle $12\times12\times8\text{ nm}^3$ induced by cosine CMP with ${f_0}$ = $17.7$ GHz, $H_\text{mw} = 0.045$ T, and optimal  $R$  for different temperature $T$ (0 and 300 K). (b)  At temperature $T=300$ K, the average magnetization switching (bold line) of nanoparticle $12\times12\times8$ $\text{nm}^3$ is shown from the 30 independent switchings induced by cosine CMP with $H_\text{mw}=0.045$, $f_0=17.7$ GHz, and $R=0.39$ GHz.  } 
    \label{fig:2}
\end{figure}

Firstly we investigate the magnetization switching of the nanoparticle $12\times12\times 8$ nm$^3$ driven by cosine CMP with the initial frequency $f_0=17.70$ GHz and the microwave field $H_\text{mw}=0.045$ T, which are similar parameters to the study \cite{islam2021fast} for $T=0$ K. By keeping $f_0=17.70$ GHz and $H_\text{mw}=0.045$ T fixed, then we try to determine the optimal $R$ (at which the switching is fastest) which is obtained $ 0.42$ GHz for $T=0$. Then we include the temperature and relax the system. Afterward, we apply the cosine CMP to the system and study the switching at finite temperatures. In principle, we simulate 30 independent switchings by varying the random numbers$\slash$thermseeds  for the same parameters of material and of cosine CMP ($f_0=17.70$ GHz, $H_\text{mw}=0.045$ T and $R=0.39$ GHz) and take the ensemble average as shown in figure \ref{fig:2} (b). In this way, we estimate each observations$\slash$data points of this study.  Now for $T=300$ K, we study the system $12\times12\times 8$ nm$^3$ by cosine CMP with the parameters ($f_0=17.70$ GHz, $H_\text{mw}=0.045$ T and $R=0.42$ GHz), but the switching is not obtained as expected which shown by the red line in the \ref{fig:2}(a). Then $R$ is optimized (for fixed $f_0=17.70$ GHz and $H_\text{mw}=0.045$ T), and for the optimal $R=0.39$ GHz, we find the fastest magnetization switching as shown by the blue line in Figure \ref{fig:2} (a). Therefore, it is noted that optimal $R$ depends on temperature, which is a crucial issue that needs to be addressed for realistic applications.

\begin{figure}
    \centering
    \includegraphics[width=0.45\textwidth]{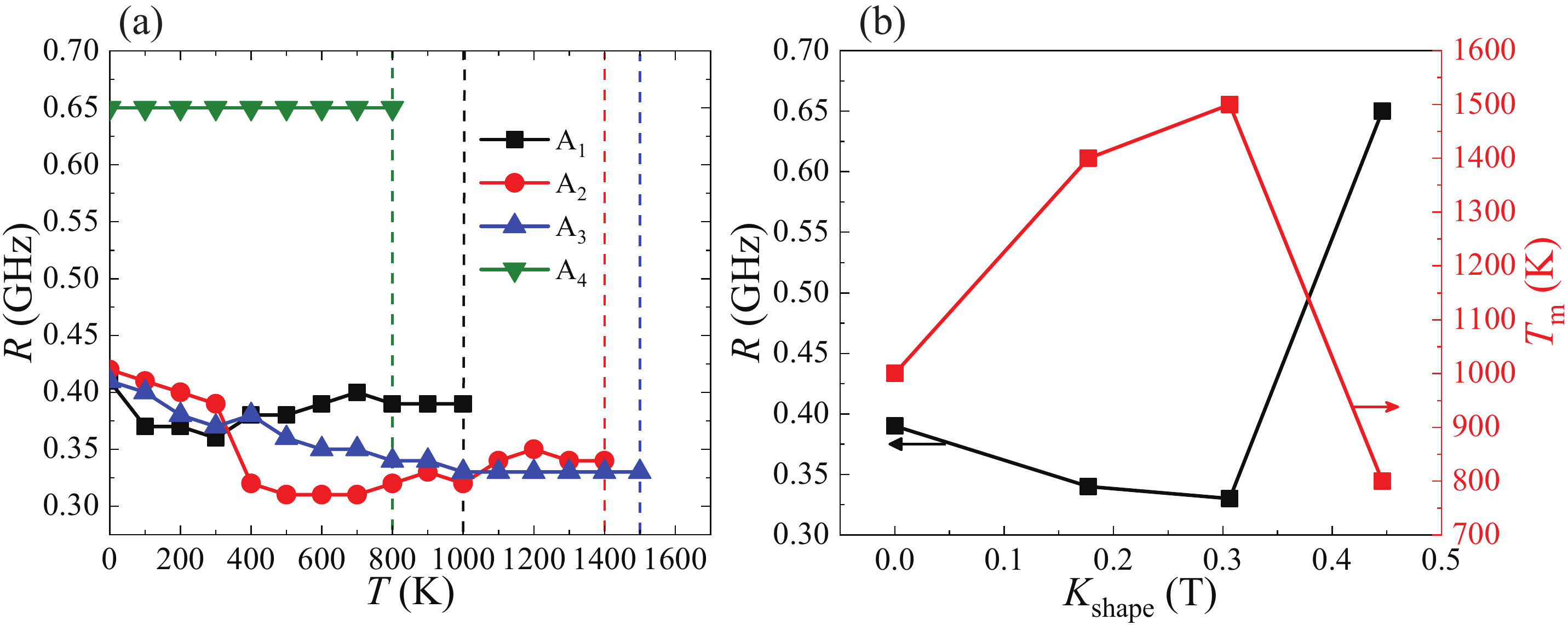}
    \caption{(a) The optimal  $R$ as a function of $T$ for different $K_\text{shape}$ ($A_1=8\times8$ nm$^2$, $A_2=12\times12$ nm$^2$, $A_3 =16\times16$ nm$^2$, and $A_4=22\times22$ nm$^2$). (b) $T_\text{m}$  (red line) and optimal $R$ corresponding to $T_\text{m}$   (Black line) are the function of shape anisotropy $K_\text{shape}$.}
    \label{fig:3}
\end{figure}

It is reported that, with enlarging the volume of the samples$\slash$nanoparticles sizes, the thermal stability ($E_a/E_t$) increases since the anisotropy energy, ($E_a=KV$, $K$ is effective anisotropy coefficient) is proportional to volume, which increases the energy barrier \cite{koche2000,lia2004,de2017,iwata2018,liu2020} but thermal energy $E_t=k_BT$ does not depend on volume. In this study, we focus on the samples with increasing cross-sectional area $A=xy$ while the thickness $h=8$ nm is kept constant so that the volume ($V=Ah$) is enlarged. Specifically, our considered samples are with $A_1=8 \times 8 $, $A_2=12 \times 12 $,  $A_3=16 \times 16 $ and $A_4=22 \times 22 $ nm$^2$ while the thickness $z=h = 8$ nm is fixed. Therefore, in these sample shapes, the demagnetization field$\slash$shape anisotropy (which is the opposite of the crystalline anisotropy) would be induced because of the magnetization of the nanoparticle. For different $A$, by finding the demagnetization-factors $N_z$ and  $N_x$ \cite{dubowik1996,aharoni1998d} analytically, the  shape anisotropy  of the samples $ {K}_\text{shape} = \mu_0 (N_z-N_x)M_\text{s} m_{z} \hat{\mathbf{z}}$ has been calculated. So, shape anisotropy $K_\text{shape}$  are $0$ T, $0.17718$ T, $0.3064$ T and $0.4459$ T, for $A_1$, $A_2$, $A_3$ and $A_4$ respectively. The $\mathbf{H}_\text{shape}$ actually opposes the anisotropy field  $\mathbf{H}_\text{ani}$ and reduces the stability as well as resonance frequency  $f_0=\frac{\gamma}{2\pi}\left[H_\text{ani} - \mu_0 (N_z-N_x)M_\text{s}\right]$. Therefore, there is an issue to study how increment of the shape anisotropy, as well as the volume of the sample, affect the thermal stability and the parameters of cosine CMP.  
Purposely, we investigate the cosine CMP-driven magnetization switching of the patterned samples of $A_1$,  $A_2$  $A_3$, and $A_4$ by varying $R$ and estimate optimal $R$ for different $T$.
Fig \ref{fig:3}(a) demonstrates the change of optimal $R$ as a function of thermal effect, $T$ for different samples. Optimal $R$ shows a decreasing trend for lower (smaller cross-sectional area) $K_\text{shape}$, albeit with some fluctuation. However, for higher $K_\text{shape}$,  optimal $R$ is larger (the magnetization switching is faster) and remains constant with $T$.    
For each sample, there is a maximal temperature $T_\text{m}$ for which the magnetization switching is valid, which is indicated by the vertical dashed line in Figure \ref{fig:3}(a). $T_\text{m}$'s are higher than the room temperature. The above findings are useful for device applications. In Fig \ref{fig:3}(b), we explicitly plot  $T_\text{m}$ and optimal $R$ at $T_\text{m}$ as a function $K_\text{shape}$, it is found that $R$ decreases till a certain value of $K_\text{shape}$  and for further increment of $K_\text{shape}$, $R$  increases significantly, i.e., the switching time becomes smallest.   The reason can be attributed as (using basic knowledge) the effective saturation magnetization $M_s$ decreases with temperature because of the spin-wave generation \cite{Kittelbook}. For $A_1$, $A_2$ and $A_3$, i.e., smaller volume, $M_s$ decreases faster with $T$ refers to the study \cite{islam2021thermally}, so optimal $R$ decreases as a function of $T$ with some fluctuations, which might be absent if one can take a large number of ensemble average, and the change of optimal $R$ with $T$ would be more consistent. However, for larger $A_4$, i.e., larger volume, the decrement of $M_s$ with $T$ is not significant, so optimal $R$ remains constant. For $A_1$, $A_2$ and $A_3$,  $T_\text{m}$ increases because of the increment of the sample volume, rather than $K_\text{shape}$ as it is not dominant, which plays a role to increase the thermal stability. But for the larger $A_4$, $K_\text{shape}$ becomes dominant and it reduces the uniaxial anisotropy as well as the energy barrier which separates the two stable states. This is why thermal stability decreases, i.e.,  $T_\text{m}$ decreases significantly.     

\begin{figure}
    \centering
    \includegraphics[width=0.45\textwidth]{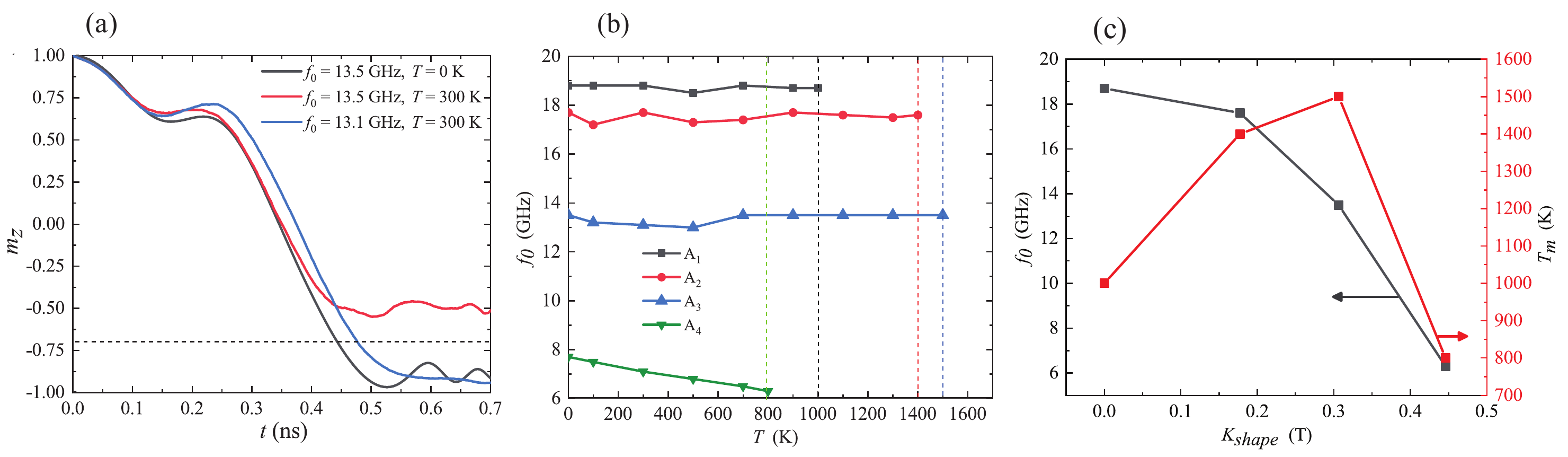}
    \caption{(a) The switching of $m_z$ of nanoparticle with $A_3 =16\times16$ nm$^2$ driven by the cosine CMP (using $H_\text{mw} = 0.045$ T and $R=0.37$ GHz) with optimal frequency ${f_0}$ in GHz for different finite temperature, $T$. (b) Temperature $T$ dependence of $f_0$ while $H_\text{mw}=0.045$ T and the optimal $R$ at corresponding finite temperature, $T$. (c) The maximal temperature $T_\text{m}$ (red line)  and $f_0$ (black line) at $T_\text{m}$  as the function of the shape anisotropy $K_\text{shape}$.}
 \label{fig:4}
\end{figure}

Subsequently, we determine the optimal $f_0$ of cosine CMP, with the investigation of magnetization switching, with $T$ by keeping amplitude $H_\text{mw}=0.045$ T and the optimal $R$ for corresponding $T$ fixed. Figure \ref{fig:4}(a) shows the magnetization switching (black line) induced by cosine CMP with $H_\text{mw}=0.045$ T, $f_0= 13.5$ GHz, and $R=0.37$ GHz for $T$= 0 K, but at $T$= 300 K, the magnetization switching (red line) is no longer valid with these parameters. Then we tune $f_0$ by keeping  $H_\text{mw}=0.045$T and optimal $R$ (at $T$= 300 K) fixed and we obtain magnetization switching (blue line) with $f_0=13.10$ GHz.
Then for different  $K_\text{shape}$  or samples, we investigate the variation of $f_0$ with $T$ and demonstrate in Figure \ref{fig:4}(b). With increasing the $K_\text{shape}$, the initial frequency $f_0$  decreases significantly, which is expected as the  $\mathbf{H}_\text{shape}$  actually acts in the opposite direction of the $\mathbf{H}_\text{ani}$ and thus $f_0$ decreases as shown in the Figure \ref{fig:4} (a). For the samples of ${K}_\text{shape}=0$ T, 0.17718 T and 0.3064 T, minimal $f_0=$ remains almost constant up to a maximal temperature $T_\text{m}$ which is useful for practical device realization. But, for the sample of ${K}_\text{shape}= 0.4459$ T, optimal shows a decreasing trend. This is because of the reduction of effective $M_s$ (since the demagnetization field is strong), which leads to a decrement of intrinsic frequency $\gamma M_s$ as well. As a result, all dynamics become slow, as though the time scale of the magnetization dynamics has been expanded. Explicitly, $T_\text{m}$ and  optimal $f_0$ at $T_\text{m}$  as a function of $K_\text{shape}$  are plotted in the Figure \ref{fig:4}(b), it is observed that $T_\text{m}$ initially increases up to certain temperature and then $T_\text{m}$ decreases rapidly for further increment of $K_\text{shape}$ . This is because the $K_\text{shape}$ opposes and reduces the effective anisotropy $H_k$ and thus reduces the energy barrier.   

\begin{figure}
\centering
   \includegraphics[width=0.45\textwidth]{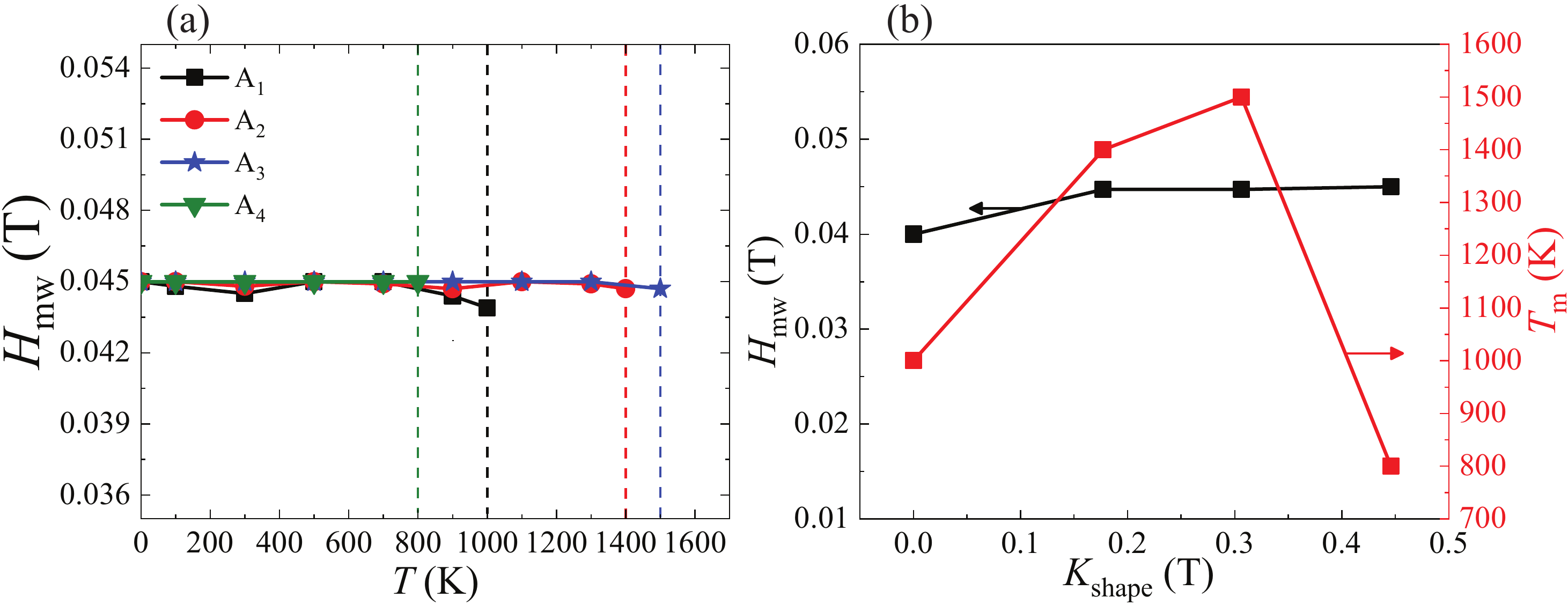}  
   \caption{(a) Temperature $T$ dependence of $H_\text{mw}$ while minimal $f_0$ and $R$ at corresponding $T$ are fixed. (b) $T_\text{m}$ (red line) and the minimal $H_\text{mw}$ (black line) at $T_\text{m}$ as a function of the shape anisotropy $K_\text{shape}$.}
   \label{fig:5}
\end{figure}

Lastly, we study how the minimally required $H_\text{mw}$ of cosine CMP varies with temperature $T$ by keeping the optimal $f_0$ and $R$ at corresponding $T$ fixed. For $\mathbf{K}_\text{shape}=0$ or $A_1=8\times8 $ nm$^2$, the variation of $H_\text{mw}$ (black line) presented in the Figure \ref{fig:5}(a) and it is found that the required $H_\text{mw}$ remains almost constant up to a maximum temperature $T_\text{m}$ which is indicated by the vertical dashed line. After $T_\text{m}$, the magnetization switching is not obtained.  Similarly, for other $K_\text{shape}$  or samples, we study the magnetization switching to find minimal $H_\text{mw}$ of cosine CMP varies with temperature $T$ by keeping the optimal $f_0$  and $R$ at corresponding $T$ fixed and presented in Figure \ref{fig:5}(a). It is noted that for higher $T$ and lower $K_\text{shape}$, the required  $H_\text{mw}$  are smaller. To be more explicit, the  $T_\text{m}$ and the minimal $H_\text{mw}$ at $T_\text{m}$ [$H_\text{mw}(T_m)$]  are plotted in Figure \ref{fig:5}(b) which shows that thermal stability $T_\text{m}$ increases with $K_\text{shape}$  up to a certain value and then decreases with the further increment of $K_\text{shape}$. It is because of a similar reason as the increment of the sample volume, rather than $K_\text{shape}$  which is not dominant, increases the thermal stability. But for the larger $A_4$, $K_\text{shape}$  becomes dominant and it reduces the uniaxial anisotropy and thus the thermal stability decreases, i.e.,  $T_\text{m}$ decreases significantly.
From the above study, we estimate the minimal $H_\text{mw}$, $f_0$, and optimal $R$ at different $T$ for different $K_\text{shape}$  or the cross-sectional areas.  the minimal $H_\text{mw}$, $f_0$ and optimal $R$ at three different $T$ (= 0 K , 300 K and  $T_\text{m}$) are summarized in the Table \ref{table-1}.

\section{Discussions and Conclusions}
The recent study \cite{islam2021fast} has demonstrated that the cosine CMP is capable of driving fast and energy-efficient magnetization switching of a nanoparticle with $T=0$. However, this study investigates the cosine CMP-driven magnetization switching of the nanoparticle by including finite $T$ since the temperature is ubiquitous in nature. We found that cosine CMP-driven fast and energy-efficient switching is still valid in finite $T$ or even at room temperature, which is important in practice.  For the lower volume of samples, the required minimal amplitude, the optimal $R$, and minimal frequency decrease with temperature because of the decrement of magnetization with $T$. We also find that with the increment of easy-plane shape-anisotropy, the required initial frequency of cosine CMP significantly reduces. For the larger volume of nanoparticles, the parameters of cosine CMP remains constant for a wide range of temperature which is useful practically.  It is mentioned that there is a very recent study which demonstrates how to generate such cosine CMP in practice \cite{sameh2023josephson} and furthermore, several technologies \cite{cai2013,cai2010} are available to generate such cosine CMP.  It is suggested to simulate magnetization switching for a real nanoparticle and thus obtain the optimal parameters; later these parameters can be employed for all the same nano-particles of the device. In addition, it is mentioned that this strategy might be applicable to switch the magnetization of synthetic antiferromagnetic$\slash$ferrimagnetic nanoparticles.  Therefore, the above findings may pave the way to realize the future-generation highly dense and speedy processing memory devices.

\section*{Acknowledgments}
M T Islam acknowledges the National Key R\&D Program of China (Grant No.
2021YFA1202200), the Khulna University Research Cell (Grant No. KU/RC-04/2000-158), Khulna, Bangladesh, and the Ministry of Education (BANBEIS, Grant No. SD2019972).

\appendix

\appendix

\section[Calculation of $\frac{dE}{dt}$]{Calculation of the rate of change of energy}
The energy is a function of magnetization and microwave field, i.e., $E(\mathbf{m}$, $\mathbf{H}_\text{tot})$. The rate of change of energy is then,

\begin{equation*}
    \frac{dE}{dt} = \frac{\partial E}{\partial \mathbf{m}}\frac{d \mathbf{m}}{dt} + \frac{\partial E}{\partial \mathbf{H}_\text{tot}}\frac{d \mathbf{H}_\text{tot}}{dt}
\end{equation*}
where, $\mathbf{H}_\text{tot}=\mathbf{H}_\text{eff}+\mathbf{H}_\text{mw}$, 
the Hamilton's equation of motion for the system is,
\begin{align*}
\frac{\partial E}{\partial \mathbf{H}_\text{tot}} = - \mathbf{m}\\
    \frac{\partial E}{\partial \mathbf{m}} = - \mathbf{H}_\text{tot}
\end{align*}

Now, after substituting the values of $\frac{d\mathbf{m}}{dt}, \frac{\partial E}{\partial \mathbf{H}_\text{tot}}$ and $\frac{\partial E}{\partial \mathbf{m}}$ in the equation, we get

\begin{align*}
    & \frac{dE}{dt} = - \mathbf{H}_\text{tot} \cdot (- \gamma \mathbf{m} \times \mathbf{H}_\text{tot} - \alpha \mathbf{m} \times (\gamma \mathbf{m} \times \mathbf{H}_\text{tot})) \\
    & \hspace{5cm} - \mathbf{m} \cdot \frac{d \mathbf{H}_\text{mw}}{dt} \\
    \Rightarrow & \frac{dE}{dt} = \gamma \mathbf{m} \cdot (\mathbf{H}_\text{tot} \times \mathbf{H}_\text{tot}) + \alpha \gamma \mathbf{H}_\text{tot} (\mathbf{m} \times (\mathbf{m} \times \mathbf{H}_\text{tot})) \\
    & \hspace{5cm} - \mathbf{m} \cdot \frac{d \mathbf{H}_\text{mw}}{dt} \\
    \Rightarrow & \frac{dE}{dt} = \alpha \gamma \mathbf{H}_\text{tot} (\mathbf{m} \times (\mathbf{m} \times \mathbf{H}_\text{tot})) - \mathbf{m} \cdot \frac{d \mathbf{H}_\text{mw}}{dt} \\
    \Rightarrow & \frac{dE}{dt} = \alpha \gamma (\mathbf{m} \times \mathbf{H}_\text{tot}) \cdot (\mathbf{H}_\text{tot} \times \mathbf{m}) - \mathbf{m} \cdot \frac{d \mathbf{H}_\text{mw}}{dt} \\
    \Rightarrow & \frac{dE}{dt} = - \alpha \gamma (\mathbf{m} \times \mathbf{H}_\text{tot}) \cdot (\mathbf{m} \times \mathbf{H}_\text{tot}) - \mathbf{m} \cdot \frac{d \mathbf{H}_\text{mw}}{dt} \\
    \Rightarrow & \frac{dE}{dt} = - \alpha \gamma \left|\mathbf{m} \times \mathbf{H}_\text{tot}\right|^2 - \mathbf{m} \cdot \frac{d \mathbf{H}_\text{mw}}{dt}
\end{align*}

We represent the second term of right-hand side of the above equation as $\dot{\mathscr{E}}$. We can write it in the explicit form as the following. The rate of change of microwave field $\mathbf{H}_\text{mw}$ is
\begin{align*}
    \dot{\mathbf{H}}_\text{mw} & = \frac{d\mathbf{H}_\text{mw}}{dt} \\
    & = \frac{d}{dt}\left(H_\text{mw} \left[ \cos\phi(t) \hat{\mathbf{x}} + \sin\phi(t) \hat{\mathbf{y}}\right] \right) \\
    & = H_\text{mw} \left[ -\sin\phi(t) \hat{\mathbf{x}} + \cos\phi(t) \hat{\mathbf{y}}\right] \frac{d\phi}{dt} \\
    & = H_\text{mw} \left[ -\sin\phi(t) \hat{\mathbf{x}} + \cos\phi(t) \hat{\mathbf{y}}\right] \\
    & \hspace{3cm} \left[ \frac{\phi(t)}{t} - \frac{d}{dt} \left(\frac{\phi(t)}{t}\right) t \right]
\end{align*}

And, the magnetization $\mathbf{m}$ is recast by
\begin{align*}
    \mathbf{m} & = m_x \hat{\mathbf{x}} + m_y \hat{\mathbf{y}} \\
    & = \sin\theta(t) \cos\phi_m(t) \hat{\mathbf{x}} + \sin\theta(t) \sin\phi_m(t) \hat{\mathbf{y}}
\end{align*}
where $\theta(t)$ and $\phi_m(t)$ is the polar and azimuthal angle of the magnetization $\mathbf{m}$ respectively.

Now, substituting the expression of  $\mathbf{m}$ and $\dot{\mathbf{H}}_\text{mw}$ in the expression of $\dot{\mathscr{E}}$, we get,
\begin{align*}
    \dot{\mathscr{E}} & = - \mathbf{m} \cdot \dot{\mathbf{H}}_\text{mw} \\
    & = H_\text{mw} \sin\theta(t) [ -\sin\phi(t) \cos\phi_m(t) \\
    & \hspace{0.5cm} + \cos\phi(t) \sin\phi_m(t)] \cdot \left[ \frac{\phi(t)}{t} - \frac{d}{dt} \left(\frac{\phi(t)}{t}\right) t \right] \\
    & = H_\text{mw} \sin\theta(t) \sin \left( \phi(t) - \phi_m(t)\right) \\
    & \hspace{3.5cm} \left[ \frac{\phi(t)}{t} - \frac{d}{dt} \left(\frac{\phi(t)}{t}\right) t \right] \\
\end{align*}

We can define $\Phi (t) = \phi_m(t) - \phi(t)$, and then we get,
\begin{align*}
    \dot{\mathscr{E}} = H_\text{mw} \sin\theta(t) \sin \Phi(t) \left[ \frac{\phi(t)}{t} - \frac{d}{dt} \left(\frac{\phi(t)}{t}\right) t \right]
\end{align*}
where $\left[ \frac{\phi(t)}{t} - \frac{d}{dt} \left(\frac{\phi(t)}{t}\right) t \right]$ defines the angular frequency $\omega(t)$.

\bibliography{bibliography}
\end{document}